\documentclass[runningheads,a4paper]{llncs}

\usepackage[labelfont=bf, figurename=Fig., tablename=Table]{caption}
\usepackage{amssymb}
\usepackage{graphicx}
\usepackage{rotating}
\usepackage{multicol}
\usepackage{multirow}
\usepackage{longtable}
\usepackage{url}
\usepackage[T1]{fontenc}
\usepackage[applemac]{inputenx}

\urldef{\mailsa}\path|mary-ann.blaetke@ovgu.de|
\newcommand{\keywords}[1]{\par\addvspace\baselineskip
\noindent\keywordname\enspace\ignorespaces#1}

\usepackage{xcolor}

\begin{document}

\mainmatter

\title{Predicting Phenotype from Genotype Through Automatically Composed Petri Nets}
\titlerunning{Automatically Composed Petri Nets}

\author{ Mary Ann Bl\"atke$^1$, Monika Heiner$^2$ and Wolfgang Marwan$^1$}
\authorrunning{Bl\"atke, Heiner, Marwan}

\institute{$^1$ Magdeburg Centre for Systems Biology and Lehrstuhl f\"ur Regulationsbiologie, Otto-von-Guericke-Universit\"at, Magdeburg, Germany\\ 
$^2$ Chair of Data Structures and Software Dependability, Brandenburg Technical University, Cottbus, Germany\\ 
\mailsa\\}

\toctitle{Lecture Notes in Computer Science}
\tocauthor{Authors' Instructions}
\maketitle

\begin{abstract}

We describe a modular modelling approach permitting curation, updating, and distributed development of modules through joined community effort overcoming the problem of keeping a combinatorially exploding number of monolithic models up to date. For this purpose, the effects of genes and their mutated alleles on downstream components are modeled by composable, metadata-containing Petri net models organized in a database with version control, accessible through a web interface. Gene modules can be coupled to protein modules through mRNA modules by specific interfaces designed for the automatic, database-assisted composition. Automatically assembled executable models may then consider cell type-specific gene expression patterns and the resulting protein concentrations. Gene modules and allelic interference modules may represent effects of gene mutation and predict their pleiotropic consequences or uncover complex genotype/phenotype relationships. Forward and reverse engineered modules are fully compatible. 



\end{abstract}
\section{Introduction}

Systems biology witnesses the evolution of experimental high-throughput me-thods with steadily increasing power regarding the quantification of nucleic acids, proteins, covalent modifications, and metabolites. Soon, these methods will broad\-ly allow \textit{omics} scale analyses of molecules involved in cellular regulatory control that capture the time-resolved response to stimulation or (genetic) network perturbation \cite{Ideker:2012kl}. These advances challenge the development of integrative and modular modelling frameworks that support the combination of findings obtained through different, qualitative and quantitative experimental approaches. To be useful, such models will need to be multi-level in terms of integrating multiple levels of abstraction in causally connecting experimentally well established processes at the molecular level at adjustable resolution of details with higher level phenomena. These may include cell fate decisions for simulating the intrinsic heterogeneity of clonal populations of cells following individual trajectories during development. Petri nets are an ideal formalism for the formal description of processes at multiple levels of abstraction for systems biology purposes \cite{Fisher:2007ai,Heiner:2009tg,Pinney:2003dz}.

For the sake of creating realistic scenarios, it will presumably become indispensable to compare, on a regular basis, simulation results on one and the same network topology as obtained by employing continuous, stochastic, and hybrid paradigms. In Snoopy, a given Petri net graph can be interpreted and simulated as continuous (ODE), stochastic, hybrid, or simply as qualitative model with export option to SBML \cite{Rohr:2010dk}. Interpretation as coloured Petri net, again in Snoopy, provides advanced options for biomodel engineering \cite{Liu:2012oq} as coloured Petri nets combine the formalism of Petri nets with the expressive power of a programming language. For this reason, and because of the intuitively accessible graphical representation, we have chosen Petri nets as framework for modelling and simulation. 

Repeated iterations of experimental data acquisition, modelling, and simulation
can evaluate the consistency of the interpretation of experimental results. This is especially true when high-throughput data come into play. However, conventional monolithic models usually represent certain aspects of a phenomenon at a certain resolution in detail and are restricted to a certain mathematical modelling paradigm. Such models can neither be easily combined with other models nor can they be easily updated by persons other than the author of the particular model. One solution to this limitation is to create a collection of Petri net modules that can be automatically linked in order to obtain and to update coherent models covering all or selected aspects of a biological process. Conceptually, such modules may be contributed, curated, and updated by individuals of the community with special expertise in certain aspects. Being organized within a database with version control, modules obtained by reverse engineering of experimental data can also be integrated and, as we will show below, help to import complex data sets into models that have been automatically generated by composition of pre-existing modules.

This paper makes three major contributions which fundamentally enhance the versatility of modular Petri net modelling by 
(1) linking regulated gene expression to protein concentration, 
(2) allowing the fully automated generation of models for the application
in genome-wide (\textit{omics}) approaches, and 
(3) linking gene mutation to complex phenotypic consequences in generating predictable models. Let us briefly elaborate on these claims.

\begin{enumerate}

\item The introduction of gene modules and mRNA modules allows to model regulated gene expression and protein biosynthesis. As the gene expression pattern of a cell is not constant and can drastically change dependent on cell type, physiological state, or experimental conditions, cells are definitely equipped with specific sets of proteins of variable relative abundance. By introducing gene modules into the model, differentially regulated gene activity and the resulting gene expression patterns directly translate into the marking of places of the protein modules. As the rates of biochemical reactions always depend on both, the kinetic rate constants and the concentrations of the reactants, changed gene expression will also change the rates of biochemical reactions, which in turn may drastically alter the dynamic behaviour of a regulatory network. Moreover, changed concentrations in regulatory proteins (e.g. transcription factors) may in a complex manner feed back onto the gene level by changing gene expression profiles. This circuitry of interwoven regulatory control becomes systematically accessible through the model. 

\item Gene and mRNA modules permit the fully automated generation of modules by simply uploading lists of gene names. This allows the automatic creation of models representing hundreds or even thousands of genes, their mRNAs and the proteins they form. By importing transcriptomic or proteomic data sets obtained in high throughput experiments \cite{r07577}, 
one can infer rate constants and reverse engineer regulatory mechanisms with the help of the model and predict changes in the proteome in response to differential gene regulation. Such models will also support the interpretation of phenomena observed in systematic RNAi screens where individual genes are knocked down \cite{p00062,p07895,p07811}.

\item Introduction of allelic influence modules extends the modelling of gene activity to the modelling of the regulatory consequences, which gene mutations have on cellular processes. This sets the formal framework to reverse engineer biomodels from complex phenotype data sets resulting from genotypic variation e.g. by employing Petri net compatible algorithms \cite{r00910,r04506,r045062}. It is obvious that such type of models have a high potential for the application to various areas from basic research to synthetic biology or personalized medicine.
 
\end{enumerate}

We are not aware of any modelling framework providing a comparable versatility and integrative power in terms of combining forward and reverse biomodel engineering. 

This paper is organised as follows.
In the next section we briefly summarise relevant own previous work on modular Petri net modelling and the automatic composition of modules with the help of a module database.
Then we will provide the rules for the generation of modules and modular models and introduce gene and mRNA modules in Section~\ref{section:pn-modules}, the application of which is demonstrated by the first case study in Section~\ref{section:case-study1}. Section~\ref{section:eukaryotes} explains the specific features of gene and mRNA modules required for modelling gene expression and its differential regulation in eukaryotes. We continue in Section~\ref{section:case-study2} with a second case study on cell differentiation and eukaryotic gene regulation define allelic influence modules and explain how these work together with gene, mRNA, and protein modules in generating models that integrate forward and reverse biomodel engineering approaches. In Section~\ref{section:conclusion} we conclude on the versatility of the approach and provide future perspectives regarding the application to synthetic biology and \textit{omics} approaches.

\section{Previous Work}
\label{section:previous-work}

Initially, we developed our modular modelling approach to represent biochemical reaction networks made of protein-protein interactions. The core idea is to make an object-oriented approach where the objects correspond to the natural modular building blocks of life. We represent individual proteins as independent and self-contained hierarchical Petri nets, called modules. Thus,  modules correspond to natural units, each of which comprises intramolecular regulatory mechanisms of the respective protein and of all its interactions with other molecules. Modules of interacting proteins can be coupled by logical nodes of identical shared subnets describing their interaction with each other. The assembly of models from a set of modules needs no further modifications at the module level. An essential advantage of this approach is the reusability of all constructed modules in models representing pathways in arbitrary combination \cite{Blatke:2012th,Blatke:2011fv}.

A crucial point of our modular modelling approach is that each module is self-contained and can be evaluated on its own. Composed networks which in terms of their performance correspond to the conventional, monolithic networks need never to be shown explicitly or even flattened. Their behaviour is entirely understandable by understanding the individual modules and the connection rules.

Our modular modelling approach has been deployed to construct modular Petri nets for two non-trivial case studies: (1) the JAK/STAT pathway in IL-6 signalling \cite{Blatke:2012th} and (2) nociceptive network in pain signalling  \cite{Blatke:2011fv}. The JAK/STAT pathway is one of the major signalling pathways in multicellular organisms controlling cell development, growth and homeostasis by regulating the gene expression. The modular network of the JAK/STAT pathway in IL-6 signalling comprises 7 protein modules (IL6, IL6-R, gp130, JAK1, STAT3, SOCS3 and SHP2). Overall, the model consists of 92 places, 102 transitions spread over 58 pages with a nesting depth of 4. The nociceptive network in pain signalling consists of several crucial signalling pathways, which are hitherto not completely revealed and understood. The latest version of the nociceptive network consists of 38 modules, among them are several membrane receptors, kinases, phosphatases and ion-channels. Thus, the model is made up by 713 places and 775 transitions spread over 325 pages with again a nesting depth of 4. 

\begin{figure}[ht!]
\centering
		\includegraphics[width=0.9\textwidth]{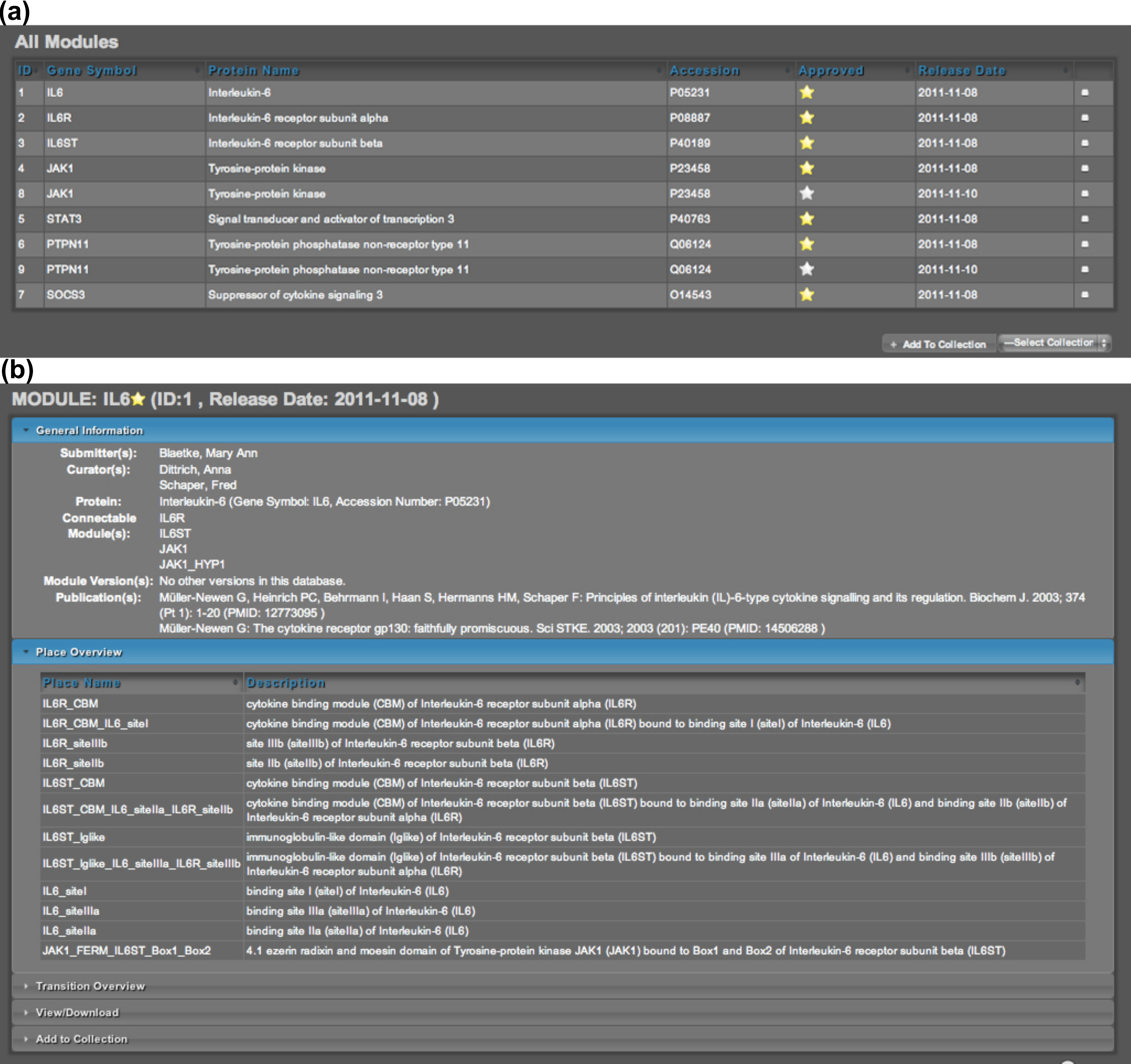}
\caption{Selected screenshots of the prototype database. (a) The web-interface enables the user to browse and/or search for modules of specific proteins. Modules can be stored in collections by selecting them. (b) Detailed information about each module can be shown on a separate page.}
\label{fig:FigDB}
\end{figure}

To support convenient module selection and network composition, we developed a database, which is accessible as prototype with a web-interface \cite{Blatke:2012th}, Figure~\ref{fig:FigDB}. 
The database holds the qualitative Petri net structure of each module, as well as the kinetic parameters mapped to each transition. 
In addition, the database contains also meta-information about the corresponding proteins (extracted from UniProt) and information about the literature used to construct the modules (extracted from PubMed), which can be linked to each module.
The organization of the modules in such a database enables the user to (1) search for individual modules, places, transitions, etc., (2) store modules in collections and (3) assemble a modular network from a chosen collection.


\section{Petri Net Modules}
\label{section:pn-modules}

\subsection{How Modules are Built and Composed}

We use Petri nets structured in the form of modules that allow the automatic composition of executable models  \cite{Blatke:2012th,Blatke:2011fv} based on Snoopy \cite{Rohr:2010dk}. The modules and their associated metadata are organized in a database accessible through a web interface designed to manage multiple versions of each module and supporting the automatic composition of models from modules (interactively) selected from a potentially big collection \cite{Blatke:2012th}. Initially, the approach was designed to model protein-protein interactions (see in Section~\ref{section:previous-work}) in the context of signal transduction networks \cite{Blatke:2012th,Blatke:2011fv}. In this paper we extend this approach by defining gene modules, mRNA modules and allelic influence modules which considerably enhances its power and versatility.

A module in general is centred around one entity describing all its interactions with other components to which the module can be linked. The entity can be a protein, a mRNA, a gene, or the specific allele of a gene. Conceptually, it is possible to extend this definition to admit other entities as well. Before going into detail, let us first briefly explain how the modular modelling approach works technically and how modules are connected to each other.

\begin{figure}[h!]
\centering
		\includegraphics[width=0.5\textwidth]{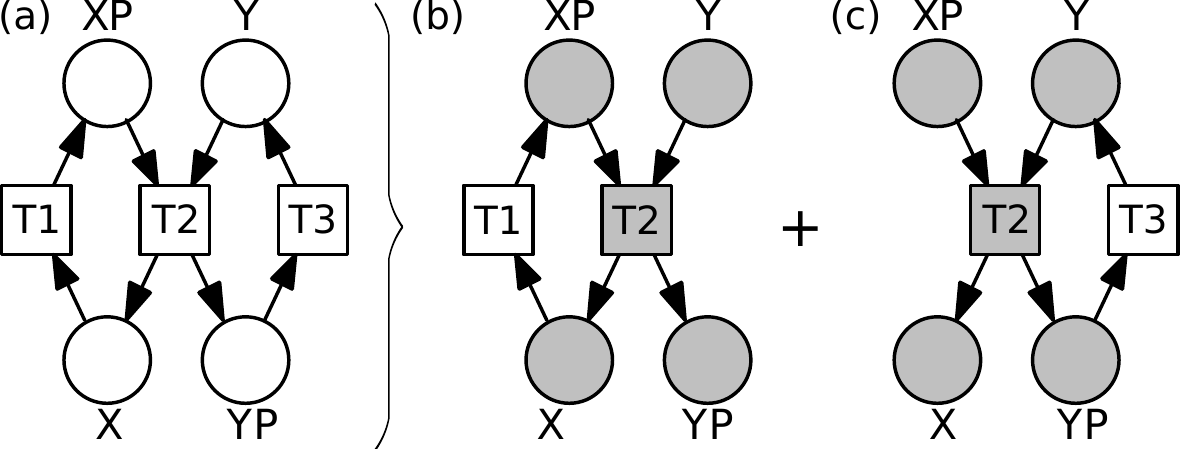}
\caption{The principle of modular modelling based on the use of logical nodes (connectors). (a) A Petri net model of the phosphorylation of protein Y by the autophosphorylating kinase X is split into modules for (b) protein X and (c) protein Y, respectively. Places and transitions that are shared by the two modules are implemented as logical nodes and shaded in grey.}
\label{fig:Fig1}
\end{figure}

The basic principle is explained taking the reaction of an autophosphorylating kinase with its substrate as example (Figure~\ref{fig:Fig1}). The kinase X autophosphorylates to give XP. XP transfers its phosphate group to the substrate Y. The phosphate group of YP then hydrolyses spontaneously. The Petri net describing these reactions (Figure~\ref{fig:Fig1}a) can be decomposed into two modules each representing the reactions of one of the two proteins involved, X and Y, respectively (Figure~\ref{fig:Fig1}b,c). Places and transitions that occur in more than one module are defined as so-called logical nodes. Logical nodes appear as multiple graphical copies of a given place or transition. In this paper, logical nodes are shaded in grey. Braking a Petri net up into modules introduces redundancy in terms of the graphical display of nodes which might appear unnecessarily complicated at first. For complex modules or when many modules are considered, the benefits are indeed tremendous as we have previously shown (\cite{Blatke:2012th,Blatke:2011fv}, and see Discussion). The biosynthesis and the degradation of a protein or a nucleic acid are modelled by separate biosynthesis or degradation modules, respectively. Accordingly, the user can choose for each protein or mRNA whether or not its biosynthesis and degradation should be considered in the synthesized model.

\vspace{-1cm}

\subsection{Definition of Modules Representing the Function of Genes}

 We now extend the initial approach constrained to protein modules by defining gene modules. A gene module considers its mechanisms of being regulated through the reversible transition between its on and off state. Assigned metadata information includes the {\it Genbank} database accession number providing the DNA sequence information as a cross-reference. Logical nodes of a gene module are used to link the gene to other modules. 

In general, multiple forms of each gene, so-called alleles, do exist that differ in one or more base pairs. These differences are due to mutations. Mutations can be silent in not altering the amino acid sequence of the encoded protein. Mutations can also be neutral in not changing the properties of the encoded protein although its amino acid sequence is changed due to mutation. These sequence polymorphisms are commonly found in wild-type populations. Alternatively, a mutation can change the properties of the encoded protein due to its altered amino acid sequence or it may even prevent the formation of the protein at all, e.g. by introducing a stop codon. Strictly following the principle of modularity in designing Petri nets, a separate gene module is created for each allele of a gene to represent mutations that change relevant properties of the gene products (RNAs and proteins) as compared to the wild-type. Entering a query for a gene, the module database will list all modules corresponding to alleles of that gene. 

We will now use two case studies to demonstrate how gene modules work. The first case  study (see Section~\ref{section:case-study1}) concerns metabolic regulation in bacteria. It is taken to explain the principle of gene modules with the help of a simple example. The second case study (see Section~\ref{section:case-study2}) considers the regulation of gene expression during the development of a eukaryotic cell. It will create a more complex scenario involving multiple layers of regulatory control. In addition, the second case study will provide an example of how modules can be obtained through reverse engineering of experimental data, it will introduce allelic influence modules and it will reveal the scalability of the approach.

\section{Case Study: The Phosphate Regulatory Network}
\label{section:case-study1}

In the first case study, we consider the response of enteric bacteria like \textit{Escherichia coli} to the limitation in inorganic phosphate which is required for the biosynthesis of nucleic acids and other cellular components (compare Figure~\ref{fig:Fig1}). 

\begin{figure}[h!]
\centering
		\includegraphics[width=1.0\textwidth]{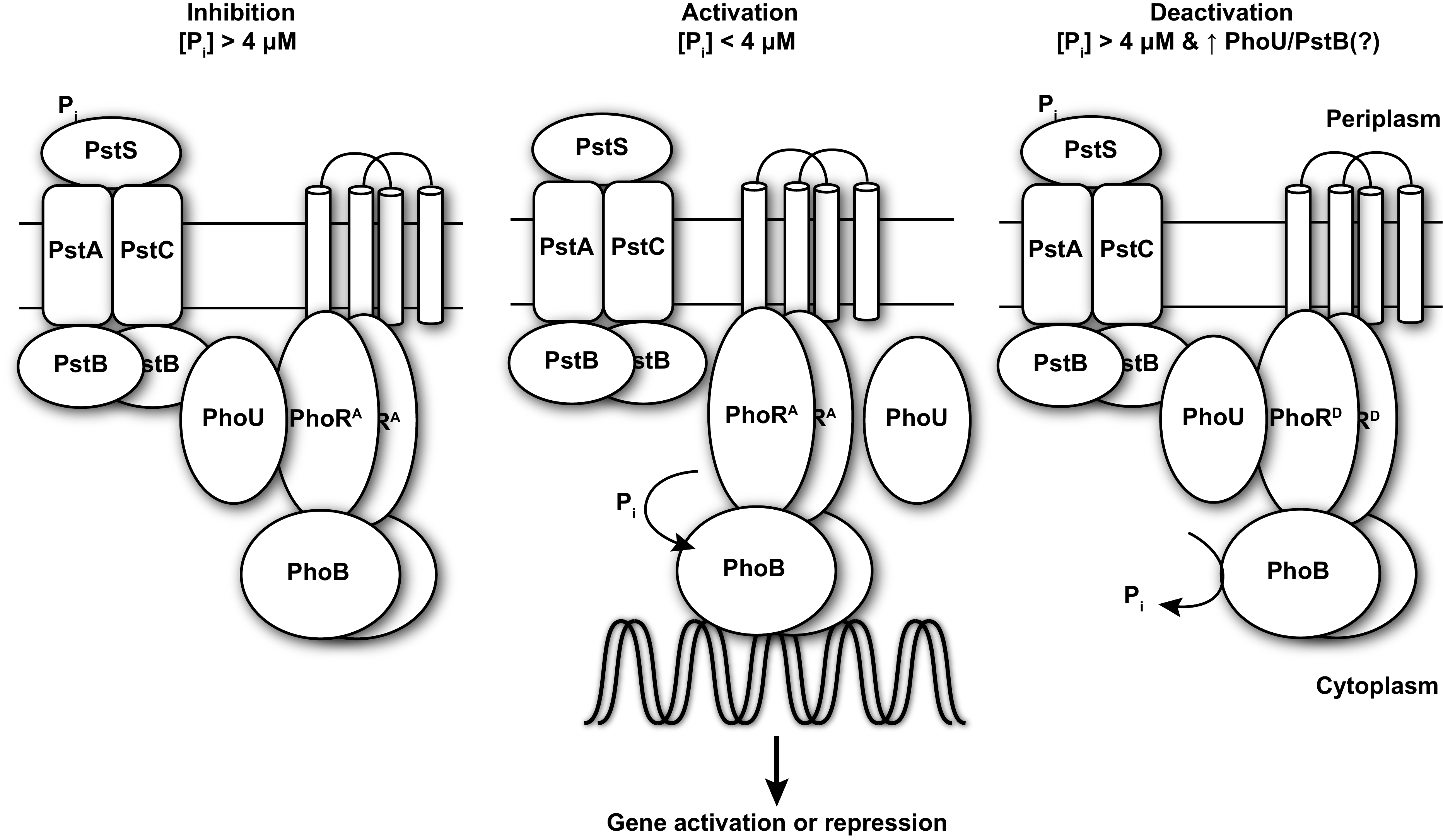}
\caption{Biochemical model for sensing extracellular inorganic phosphate (P$_i$) and transduction of the signal to control gene expression. The PstSCAB transmembrane complex serves as an ABC transporter for the uptake of environmental P$_i$. At high extracellular P$_i$ concentration, the binding protein PstS is fully saturated and this signal is relayed to the cytoplasmic part of the receptor that forms an inhibitory complex with a second transmembrane protein, the PhoR kinase via the cytoplasmic protein PhoU. If P$_i$ is low, the complex dissociates and the autophosphorylating kinase PhoR phosphorylates PhoB which, in its phosphorylated form, binds DNA to induce gene expression. When Pi subsequently increases, the compex with PhoU is formed again and PhoBP is dephosphorylated. The figure was redrawn from Hsieh and Wanner \cite{Hsieh:2010sp}.}
\label{fig:Fig2}
\end{figure}

When inorganic phosphate (P$_i$) in the environment becomes low, it may turn into a growth-limiting factor even if sufficient nutrients are available \cite{Neidhardt:1990fu}. When present, inorganic phosphate is taken up from outside of the cell through an ABC transporter system, the PstSCAB transmembrane complex (Figure~\ref{fig:Fig2} \cite{Hsieh:2010sp}). With sufficient P$_i$ outside, the PstSCAB complex actively pumps P$_i$ across the cell membrane into the cytoplasm. Under this condition, the PhoU protein, according to the proposed mechanistic model \cite{Hsieh:2010sp}, forms a complex with the pstSACB transporter system and the PhoR histidine kinase. Complex formation prevents the kinase to autophosphorylate caused by the interaction with PhoU. PhoU is a chaperone-like PhoR/PhoB inhibitory protein. When external P$_i$ is low and the PstSCAB complex is inactive, PhoU dissociates and allows the autophosphorylation of PhoR. PhoR then phosphorylates and thereby activates the transcription factor PhoB. The phosphorylated form of PhoB, namely PhoBP, then activates the transcription of at least 31 genes organised into 9 transcriptional units (\textit{eda}, \textit{phnCDEFGHIJKLMNOP}, \textit{phoA}, \textit{phoBR}, \textit{phoE}, \textit{phoH}, \textit{psiE}, \textit{pstSCAB-phoU}, and \textit{ugpBAECQ}) \cite{Hsieh:2010sp}. One of the activated genes, \textit{phoA}, encodes the PhoA protein which is a bacterial alkaline phosphatase. PhoA is exported across the membrane into the periplasm where it degrades organic phosphorous compounds to liberate P$_i$ which is then taken up into the cell to overcome the limitation. When enough P$_i$ has been formed, this system is switched off again and PhoBP is dephosphorylated.

In \cite{Marwan:2012pb}, we gave a monolithic Petri net of a simplified version of the phosphate regulatory circuitry. To cover the entire functionality we now construct
a modular Petri net model which is composed of three types of modules: (1) one protein module representing the reactions of the PhoB protein, (2) gene modules representing the regulated genes, and (3) mRNA modules representing the transcription of the gene, the translation of mRNA into the protein, and the degradation of the mRNA. The degradation of the encoded proteins is represented by degradation modules which have been introduced previously \cite{Blatke:2012th}.

\subsection{The PhoB Module}

\begin{figure}[h!]
\centering
		\includegraphics[width=0.8\textwidth]{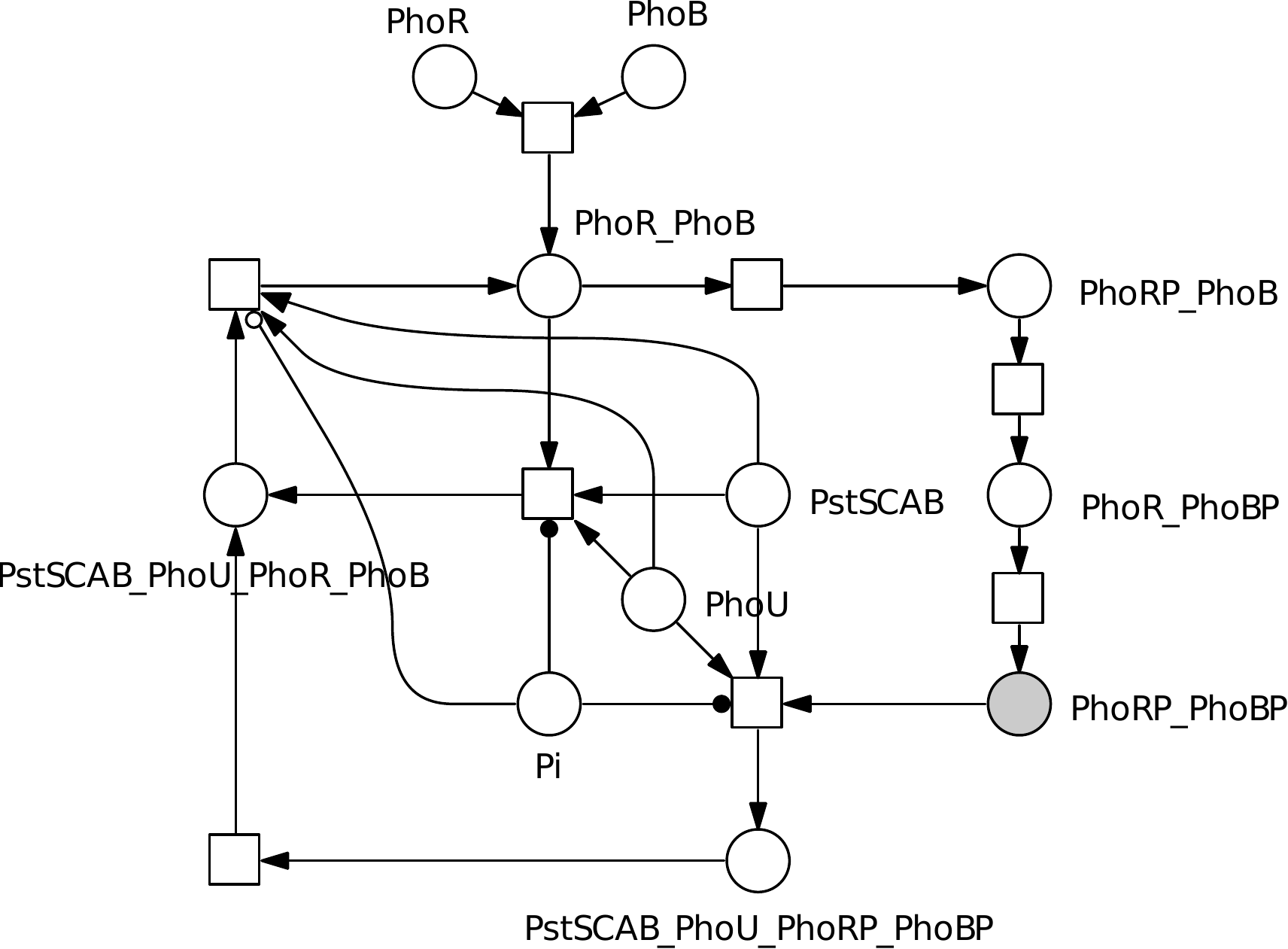}
\caption{The PhoB module representing the biochemical interactions of the PhoB protein with its binding partners PhoR, PhoU, the PstSCAB complex, and inorganic phosphate (P$_i$). PhoRP and PhoBP denote the phosphorylated forms of PhoR and PhoB, respectively.}
\label{fig:Fig3}
\end{figure}

The PhoB module (Figure~\ref{fig:Fig3}) represents the reactions of the PhoB protein in its phosphorylated (PhoBP) and dephosphorylated (PhoB) states. It also represents the complex formation of PhoB with its regulatory proteins as well as the binding of PhoB to regulatory sequences in the DNA. 

The PhoB module depicts the regulatory mechanism schematically shown in Figure~\ref{fig:Fig1}. Binding and dissociation of PhoBP in its complex with PhoRP is represented separately for each transcriptional unit. Displaying the binding of PhoBP to each regulatory site on the DNA separately keeps the Petri net graph clearly arranged and allows to reuse the structural motif of binding and unbinding reactions via copy/paste for the various regulated transcriptional units. Accordingly, PhoRP$\_$PhoBP is declared as logical place. 

To save space in Figure~\ref{fig:Fig3}, we only show binding of PhoBP to one of the nine transcriptional units.

\subsection{The Gene Modules}

\begin{figure}[ht!]
\centering
		\includegraphics[width=0.3\textwidth]{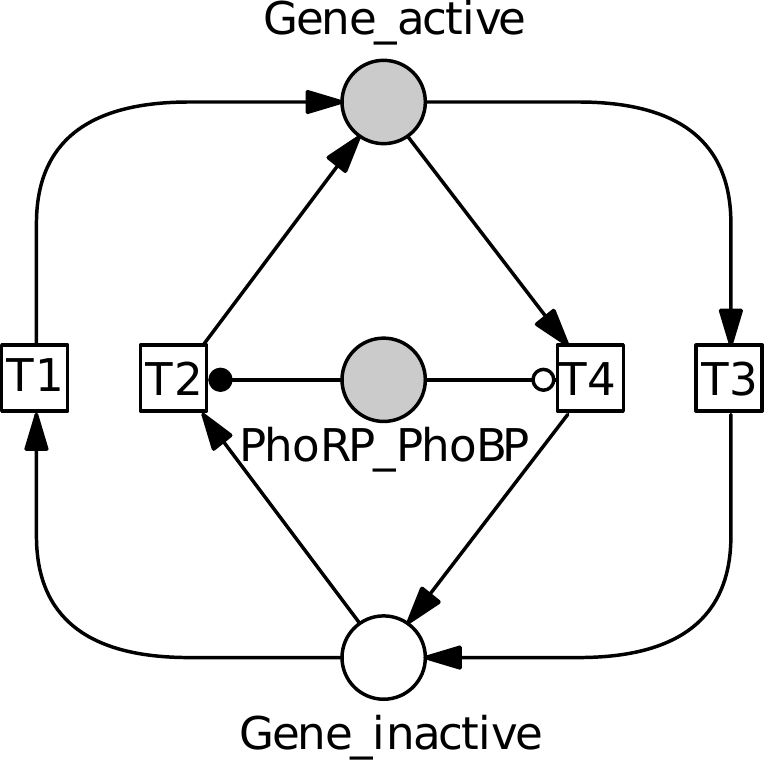}
\caption{Prototype of a gene module. The module controls the activation of any gene responsive to the phosphorylated PhoB protein. Binding of PhoRP$\_$PhoBP to the promotor renders the gene active. These regulatory interactions are modeled through a test arc activating transition T2 and an inhibitory arc blocking T4. The basal activity of the gene in the absence of PhoRP$\_$PhoBP is maintained by T1 and T3.
}
\label{fig:Fig4}
\end{figure}

A gene module represents the regulation of the gene by other factors (e.g. transcription factors) through the transition between its on and off states (Figure~\ref{fig:Fig4}). In the quantitative interpretation of the Petri net (as continuous or stochastic Petri net) the regulatory factors influence the equilibrium between the on and the off state of the gene. The on state means that the gene is transcriptionally active and that mRNA molecules can be accordingly formed  as modeled in mRNA module. 

\begin{figure}[ht!]
\centering
		\includegraphics[width=0.7\textwidth]{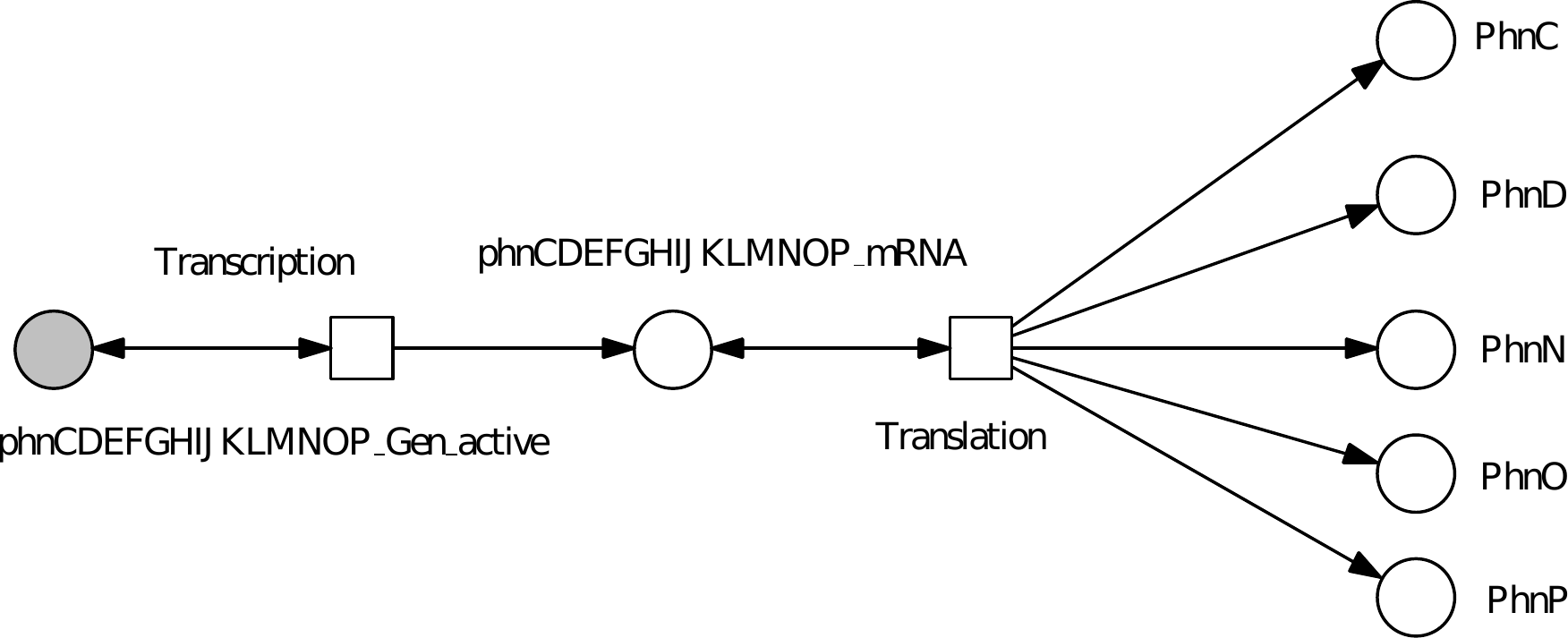}
\caption{mRNA module modelling the formation of the polycistronic message synthesized by transcription of the active \textit{phnC-P} gene. The mRNA is translated into the proteins PhnC to PhnP. For simplicity, the places for only five of the 14 proteins that are formed are shown.}
\label{fig:Fig5}
\end{figure}

In prokaryotes, several, functionally related genes can be organized into a single regulatory and transcriptional unit, a so-called operon. The genes of an operon are transcribed in the form of a single mRNA molecule, called a polycistronic message, which may encode several proteins at once. In the case of the \textit{phnCDEFGHIJKLMNOP} transcriptional unit (Figure~\ref{fig:Fig5}), one polycistronic message encoding 14 different proteins (PhoC to PhoP) is formed upon the initiation of transcription. The probability per unit of time for the initiation of transcription to occur depends on binding of PhoRP$\_$PhoBP to the regulatory region of the \textit{phnCDEFGHIJKLMNOP} operon on the DNA. From the biological point of view, polycistronic messages provide a simple mechanism for co-regulation of genes encoding proteins that work together in a cellular process. In order to obtain a model including the transcriptional regulation of all 31 genes organized into the 9 transcriptional units that are under control of the PhoB protein, 9 gene modules (\textit{phoA}, \textit{phoBR}, \textit{phoE} etc.) of analogous structure (as shown in Figure~\ref{fig:Fig4}) are required. For large scale modelling approaches, these network structures could be generated automatically or semi-automatically.

\subsection{The mRNA  Modules}

The mRNA modules depict the reactions of the respective mRNA species, namely its biosynthesis by transcription, its degradation, and its translation into the encoded proteins (Figure~\ref{fig:Fig5}). A mRNA module may in addition represent the binding of regulatory proteins to the RNA, the binding of antisense RNA influencing the stability of the message, or the processing (e.g. splicing) of the transcript, as it occurs in eukaryotes (shown in Figure~\ref{fig:Fig7} and discussed in Section~\ref{section:conclusion}.2). As transcription of the bacterial  \textit{phnCDEFGHIJKLMNOP} transcriptional unit leads to the formation of a polycistronic message, 14 proteins (PhoC to PhoP) are synthesized. The reactions (e.g. the catalytic activity) of the encoded proteins might then be considered in separate protein modules. 

\section{Modelling Eukaryotic Gene Regulation with Gene and mRNA Modules}
\label{section:eukaryotes}

\subsection{Eukaryotic Gene Modules}

The gene modules employed for modelling the regulation of eukaryotic genes are very similar to the models of prokaryotic genes as presented in the previous section for the phosphate regulatory network.  However, the regulation of eukaryotic genes is typically more complex than in prokaryotes as more protein factors and more binding sites for regulatory proteins on the DNA may be involved. All these factors together may control the on state of a gene. 

Gene regulation may involve protein binding sites on the DNA functioning as enhancers or silencers that are located several thousand base pairs distant from the genes they regulate. Proteins bound to these sites may influence the probability for transcription to be initiated through physical interactions with the protein complexes bound to the promotor of the regulated gene. These regulatory sites and the binding of regulatory proteins to these sites are represented as part of the gene module. 

\begin{figure}[ht!]
\centering
		\includegraphics[width=0.8\textwidth]{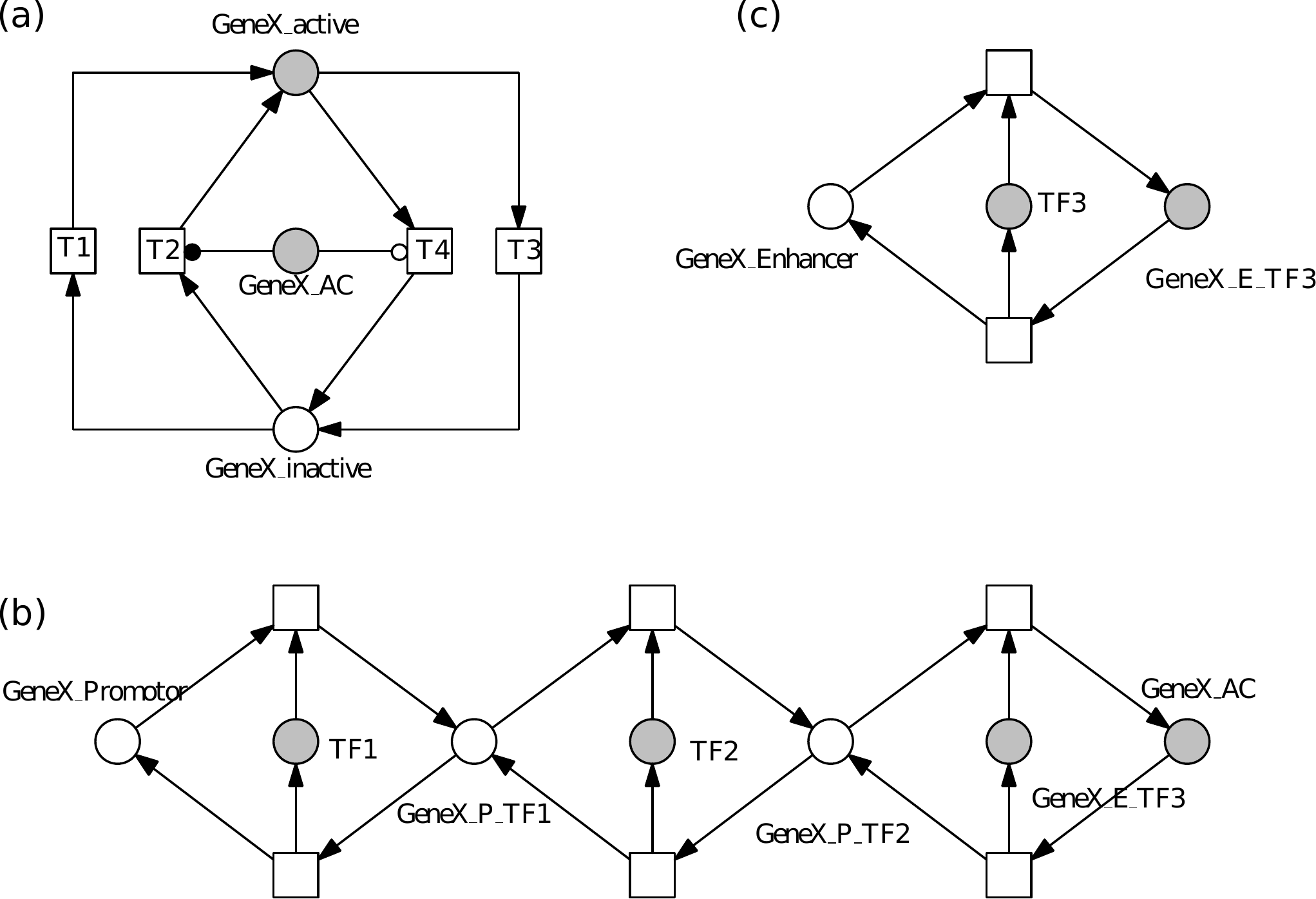}
\caption{Prototype of a eurkaryotic gene module. According to the model, the regulation of the eukaryotic gene (a) depends on more protein factors than the regulation of the prokaryotic gene shown in Figure~\ref{fig:Fig4} does. By binding to the promotor region of the gene, these factors form a multimeric protein complex, as shown for transcription factors TF1 and TF2, both of which directly bind to the promotor (b). The third transcription factor, TF3, binds first to an enhancer sequence distant from the promotor (c) and subsequently can bind to the TF1-TF2 complex to form the gene activating complex GeneX$\_$AC (b) which switches the gene into its transcriptionally active state (a). As in Figure~\ref{fig:Fig4}, gene activation by protein binding to the promotor is modeled by control arcs and the basal level of gene activity occurs through firing of transitions T1 and T3 (a). According to the definition, the gene module displays all direct molecular interactions of GeneX with the proteins that bind to its regulatory sequences. Note that binding of regulatory proteins may involve cooperative mechanisms which would be represented in the context of the gene module accordingly.}
\label{fig:Fig6}
\end{figure}

A prototype of a module representing the regulatory control of a eukaryotic gene is shown in Figure~\ref{fig:Fig6}. Making regulatory sites part of the gene module provides the advantage that potentially cooperative effects in protein binding and gene activation can be considered as part of the module. 

\subsection{Eukaryotic mRNA Modules}

\begin{figure}
\centering
		\includegraphics[width=0.9\textwidth]{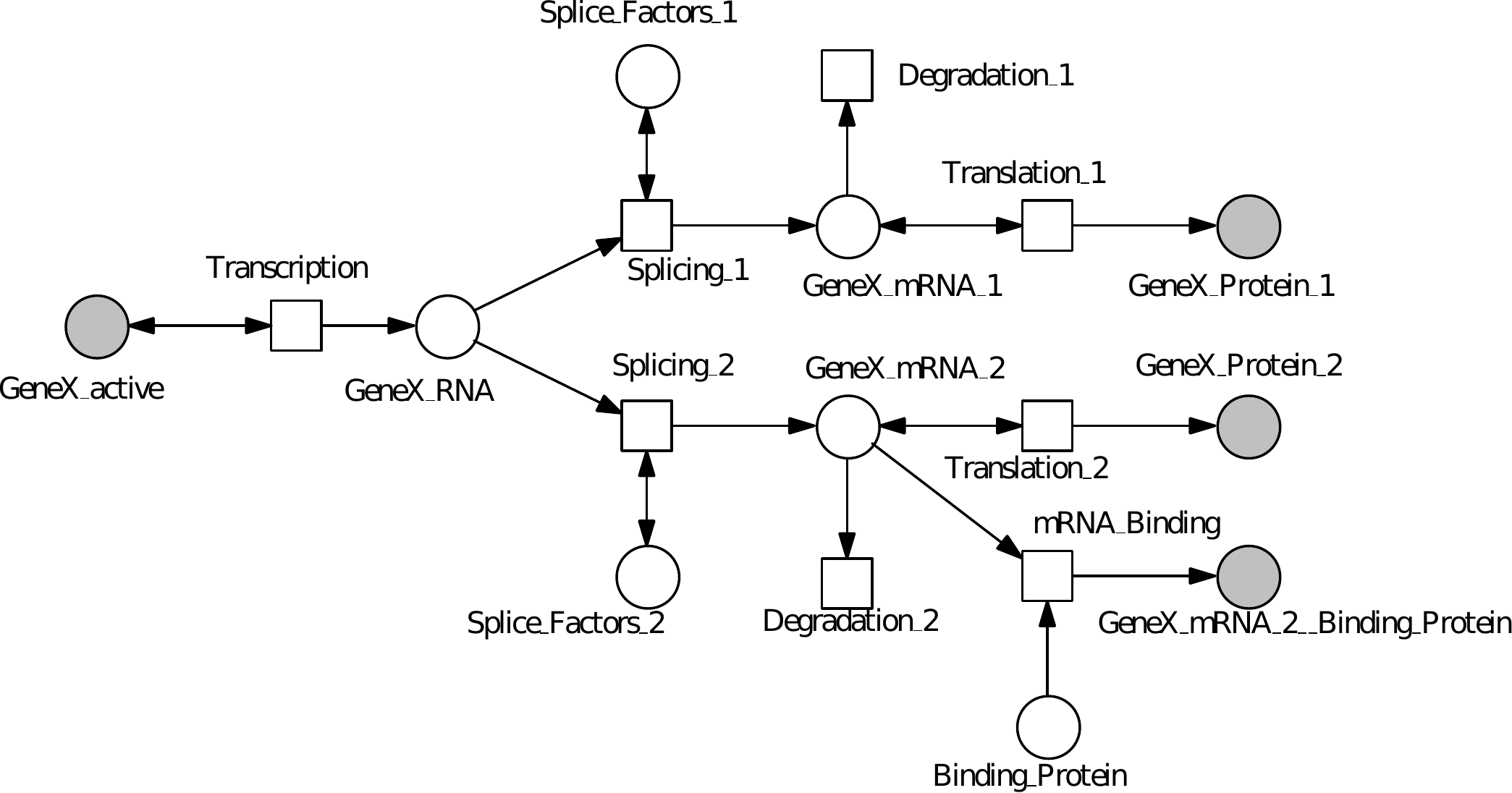}
\caption{Prototype of an eukaryotic mRNA module. Transcription of the active Gene (GeneX) leads to the formation of a primary transcript which is processed. In the example shown, the primary transcript is spliced into two alternative mRNAs, GeneX$\_$mRNA$\_$1 and GeneX$\_$mRNA$\_$2, respectively that are subsequently translated into the corresponding proteins. The mature mRNAs may bind to RNA binding proteins that regulate the stability of the mRNA and its availability for the translational maschinery. For simplicity, this reaction is shown for one of the two mRNAs only.}
\label{fig:Fig7}
\end{figure}

In addition to the biosynthesis of proteins there are several RNA-dependent processes that may be of regulatory importance especially in eukaryotic cells. Typically, the occurrence of these mechanisms depends on the considered RNA species and may also depend on physiological conditions as well as on developmental states. Each of the mechanisms described in the following has been implemented in a basic form in the mRNA module prototype shown in Figure~\ref{fig:Fig7}. 

\subsubsection{Alternative splicing.}
 Primary transcripts in eukaryotes are processed during the maturation of the final, protein-encoding mRNA. Processing includes the splicing of the RNA where non-coding introns are excised from the primary transcripts. Due to the occurrence of alternative splicing sites, it may be that differently spliced mRNAs are formed from one and the same primary transcript giving rise to proteins of partially different amino acid sequence. This splicing depends on protein factors that may be regulated depending on the physiological or developmental state of the cell. When necessary, the biochemical reactions of these slicing factors (like posttranslational modification or protein-protein interaction) may be represented in the context of protein modules with the help of logical places.
 
\subsubsection{RNA-binding proteins.}
 The half-life of mRNA species may vary between minutes and months. This variation may have different reasons in addition to the specific secondary structure of the RNA. One mechanism influencing the half-life of a given mRNA species is the binding to specific RNA-binding proteins that may store or degrade the RNA. Being bound to an RNA-binding protein like Pumilio for example, the mRNAs can be stored in the cell while being translationally inactive. Upon release from the RNA-binding proteins the mRNA may suddenly become translationally active and hence become available at relatively high concentration \cite{Gerber:2006wd}. 
 
\subsubsection{RNA interference.}
RNA interference is a natural mechanisms for the specific inactivation of the expression of eukaryotic genes, e.g. by small interfering RNAs that bind to the target RNA. The degradation of the RNAs accordingly depends on specific protein factors and on the availability of the interfering RNA \cite{Cerutti:2006bs,Uhlmann:2012hc}. 

These mechanisms may have to be considered in the context of mRNA modules and receive regulatory input from specific cellular proteins.

\subsection{Why is the Integration of Bottom-up and Top-down Models Essential, Especially in Eukaryotes?}

In some respects systems biology appears to be more difficult for eukaryotic as compared to prokaryotic cells. This may in part be due to the occurrence of fundamentally different regulatory processes in the two domains of life with non-obvious consequences of certain eukaryotic regulatory phenomena. 

There is indeed highly detailed knowledge on the canonic pathways of eukaryotic signal transduction which allows the formulation of well-structured bottom-up models representing the biochemical interactions of regulatory components like e.g. the MAP kinase cascade. Such models can be tremendously useful in understanding mechanisms in health and disease \cite{Kolch:2005qa,Orton:2009qe,Sturm:2010kl} and in finding new and powerful drugs. 

However, it is also true that many experimental findings on canonic pathway components seem to be contradictory. This may in part be due to the fact that gene expression patterns in different experimental systems and under different physiological conditions are different leading to a different composition of biochemical reactants within the cell at a given time or within a particular experiment. 

This certainly restricts the current practical value of bottom-up models without disclaiming their general usefulness. When a certain cellular process as the differentiation phenomenon in a eukaryotic cell or the progression through the cell cycle for example is to be rigorously analysed at the transcriptomic or the proteomic level, the changes in many perhaps in most of the observed components and their consequences cannot be explained on the basis of the established bottom-up models of the canonic pathways. This suggests that there are tremendous gaps in our current understanding. 

On the other hand, it seems currently impossible that the thousands of components accessible through \textit{omics} approaches can be analysed with such experimental effort as invested for the exploration of canonic pathway components. Therefore it seems self-evident that \textit{omics} data are used to rigorously reverse engineer models. A next and essential step then is to integrate these top-down with relevant bottom-up models to obtain integrated models with predictive power.


\section{Case Study: The Sporulation Control Network in \textit{Physarum polycephalum}}\label{section:case-study2}

\subsection{Sporulation is Controlled by a Gene Regulatory Network}

\textit{Physarum polycephalum} is a unicellular eukaryote belonging to the amoebozoa group of organisms  \cite{Baldauf:1997fy,Baldauf:2000eu,Glockner:2008nx}. During its relatively complex and branched life cycle, \textit{Physarum} develops into various cell types. These cell types occur in temporal order and differ in morphology (shape), molecular composition, and physiological function (\cite{Burland:1993la} and references therein). 

One of these cell types is the plasmodium, a multinucleate macroscopic single cell. Differentiation of the plasmodium can be easily studied under lab conditions as the response can be experimentally triggered by applying a brief pulse of far-red light. The light pulse sets a defined starting point on the time axis on which subsequent events are observed. During about 18 hours after the trigger, the entire plasmodial cell is extensively remodeled and fruiting bodies are formed that give rise to mononucleate haploid spores that are precursor cells of amoebal gamets which will develop at later stages of the cycle \cite{Burland:1993la}. This process is called sporulation. Please note that sporulation in bacteria, although the name is the same, is biochemically a completely different process in eukaryotic cells. 

Five to six hours after an inductive far-red pulse, the cell is irreversibly committed to sporulation. The associated molecular events are of scientific interest as the plasmodium loses some stem cell-like capabilities during commitment. The expression pattern of hundreds of genes changes \cite{Glockner:2008nx,Burland:1993la}. These changes in gene expression that normally occur can be compared to the changes that are seen in mutant cells that have lost their ability to be committed to sporulation  \cite{Barrantes:2010rw,Hoffmann:2012dp}. 

Genetic dissection of gene regulatory networks by generating mutants that are altered in the regulatory control and by analysing the phenotype which the mutation produces is a widely-used method in biosciences. Mutants may be produced through forward and/or reverse genetic approaches. In forward genetics, randomly mutated cells or organisms are screened for phenotypes of interest and the gene which causes the phenotype is identified subsequently. In the reverse genetic approach, a gene of interest is mutated and the phenotypic consequence of the generated mutation is analysed. Mutation of a gene both, in forward and reverse genetics, may either cause the loss of a protein or a change in its function. Mutation may change the activity of a protein (up or down) or it may change the specificity of its catalytic activity. In many cases, the molecular mechanisms of how a given mutation translates into the observed phenotype remain unknown for a number of years. Despite this ambiguity, the genetic approach is powerful as it rigorously establishes causal dependencies within the living organism. In most cases biochemistry alone could not fulfill this task. 

A powerful way to employ genotype/phenotype relationships for modelling and simulation is the reverse engineering of genetic data. Reverse engineering of gene expression data then provides a direct link to bottom-up models of protein-protein interactions. By reverse engineering one can establish effects, which a mutated gene (the allele of a gene) exerts on a cellular process. To represent these influences we define allelic influence modules. We will now show how allelic influence modules are built and how they are useful for the integration of top-down and bottom-up model parts into a coherent model.

\subsection{Linking Genotype to Phenotype: Allelic Influence Modules}

As gene modules, allelic influence modules are centred around the allele of a given gene. However, allelic influence modules differ from gene modules in representing the regulatory influences exerted by the allele on cellular processes by controlling the firing activities of respective transitions through read or inhibitory arcs. In reality, these influences can be rather indirect by involving numerous other, potentially unknown components. Accordingly, the allelic influence module may represent the control of molecular events like the biosynthesis of RNA by transcription or even more complex processes of potentially unknown molecular mechanism as inferred from functional studies. To make this more clear, let us consider the case study.

\begin{figure}[ht!]
\centering
		\includegraphics[width=0.5\textwidth]{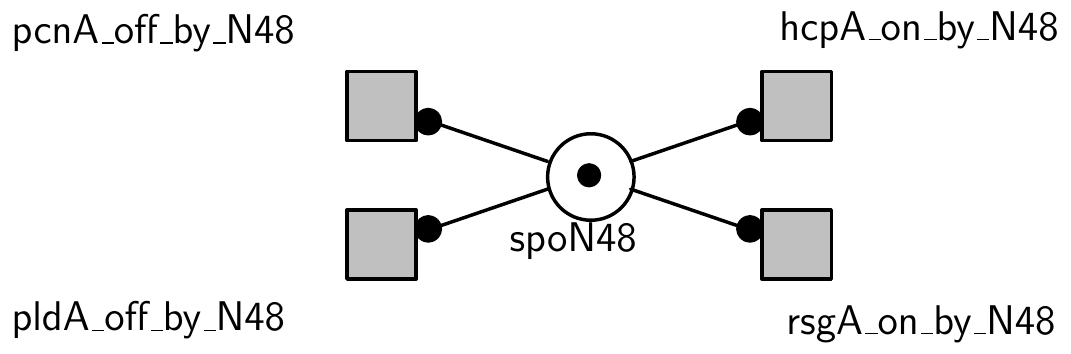}
\caption{Allelic influence module. The module represents the differential regulation of four genes, \textit{pcnA}, \textit{pldA} (down regulated), \textit{hcpA}, and \textit{rsgA} (up regulated) by the \textit{spoN48} allele of the \textit{spoN} gene in \textit{Physarum polycephalum}. The logic transitions shown here are also part of the gene modules of the four differentially regulated genes (not shown). Genes and names of their orthologs in the {\it Swissprot} database: \textit{pcnA}, proliferating cell nuclear antigen; \textit{pldA}, Phosphatidylinositol-glycan-specific phospholipase D; \textit{hcpA}, Histone chaperone ASF1A; \textit{rsgA}, Regulator of G-protein signalling 2.}
\label{fig:Fig8}
\end{figure}

In response to far-red light, \textit{Physarum} plasmodia differentially express a large number of genes several hours before the cell is irreversible committed to sporulation \cite{Hoffmann:2012dp}. Genes with both, up- and down-regulated RNAs have been identified at a genomic scale \cite{Glockner:2008nx,Barrantes:2010rw}, and the precise expression kinetics of some of them have been investigated in detail \cite{Hoffmann:2012dp}. Currently, we do neither know the molecular mechanisms through which these genes are controlled nor do we know which causal consequences the change in expression level in detail have. However, the majority of the differentially expressed genes encodes proteins with high sequence similarity to proteins of important regulatory function. 

The influence of far-red light on the expression of some of these genes in normal (wild-type) cells is shown in the Petri net of Figure~\ref{fig:Fig7}. The mRNAs that have been quantified experimentally \cite{Hoffmann:2012dp} are represented by corresponding mRNA modules (not shown). 

We have genetically dissected the underlying regulatory network with the help of mutants that are altered in the photocontrol of sporulation, as isolated in phenotypic screens (\cite{Sujatha:2005ij}; Lehmann et al., unpublished results). Representatives of one class of these mutants have lost their ability to be committed to sporulation and remain in a proliferative state throughout. Although these mutants do not respond to far-red light by sporulation, they clearly respond at the transcriptional level. However the pattern of genes that are differentially expressed in response to the stimulus significantly differs in the mutants as compared to the wild-type and also differs between mutants. The altered gene expression patterns clearly reveal the regulatory influence of the mutated genes and can be used to infer the network of regulatory control in which the different regulators inactivated in each of the mutants are interwoven. The changed pattern of differentially expressed genes can be used to reverse engineer the regulatory influence of the gene mutations. In Figure~\ref{fig:Fig8} the allelic influence module obtained for the phoN48 allele of the phoN mutant gene is shown. The proteins encoded by some of the differentially expressed genes have well-known biochemical functions in the developmental control of eukaryotic cells. In a bottom-up approach, these genes can be linked to corresponding protein modules in terms of changed protein concentrations as predicted by the model. 
 
\section{Versatility of the Approach and Future Perspectives}\label{section:conclusion}

We have described a strictly modular approach to Petri net modelling based on clearly defined types of modules corresponding to different types of molecular entities around which each module is centred: genes, RNAs, and proteins. The small number of module types and the few easy-to-follow rules for creating a module encourage community efforts in creating a collection of modules in analogy to how Wikipedia collects pages. In Petri net modules, cross references to other modules allow the automatic composition of large models that are directly executable \cite{Blatke:2012th}. A web-accessible database was constructed to manage different versions of each module and is available as a prototype \cite{Blatke:2012th}. It allows that different opinions on molecular mechanisms directly translate into alternative computational models that predict experimental findings. Moreover, working with modules provides several options for the engineering of biomodels and their scalable application to systems and synthetic biology.

\subsection{Regulatory Interactions Appear in Clear Graphical Structure}

Because each module summarizes all functional interactions of a given molecular component including its influences on other components, even complex regulatory interactions can be always displayed in the form of an easy to perceive graphical layout. Certainly, not all of the functional interactions that appear in a module necessarily have to be part of a composed model. With the support of a database, modules can be selected according to user-defined criteria and then automatically linked to give a functional model. Those interactions that do not find a counterpart in one of the selected modules remain inactive because the respective places remain unmarked in the composed Petri net.

\subsection{Modules May be Added, Removed, or Exchanged: \textit{in silico} Mutation of Networks}

The modular structure allows to remove or exchange modules when composing a model automatically without touching or even considering modules that remain unchanged. This is a considerable advantage as compared to the manual re-engineering of monolithic models which requires careful consideration of how the components are wired up with each other since overlooking connections might accidentally introduce modelling errors. With the modular approach and the built-in version control of the database, different versions of a given module can be easily exchanged. This can be very helpful to analyse how alternative kinetic mechanisms of molecular interactions would influence the overall behaviour of the system. 

For example for the phosphate regulatory network, one might wish to analyse whether or not different activation mechanisms of the PhoB protein would change the gene expression response and the performance of the feed-back loop. When working with really complex models, the modeller investigating a local mechanism has not to care about the inner life of all the numerous modules, in analogy to programming where the procedures of an approved library of subroutines do not have to be reconsidered each time they are used for building a new program. 

At the moment where the database will contain a relatively high number of modules, automatically generated models might reveal nonobvious regulatory interactions of molecular components, bring them into a quantitative context and predict nonobvious and eventually counterintuitive consequences of network activation or perturbation. This will be definitely the case when regulatory interactions at a genome-wide scale change gene expression levels that translate into an updated marking of the places of protein modules due to the change in cellular concentration of perhaps many proteins. Even without modelling, this is already obvious by just looking at the gene expression data mentioned in the \textit{Physarum} case study.

By simply removing modules from a model, one might systematically probe components for their global role in the biomolecular system. This is the \textit{in silico} complement to systematic mutant screens that are performed in genetic model organisms. \textit{In silico} mutational studies may turn out to be of great benefit for synthetic biology in all cases where systematic mutant screens cannot be applied for what reasons soever.

\subsection{Modules Can be Automatically Generated at Large Scale}

For genome scale models where the regulatory control of hundreds or thousands of genes or proteins is to be considered, automatic generation of models becomes an issue. By creating multiple copies of the module prototypes described here, modules can be generated fully automatically simply by assigning names to places and transitions. This is especially straightforward for the gene and mRNA modules but also for protein degradation modules. We propose that automatically generated large scale Petri nets will transmute into helpful tools for the reverse engineering of models from transcriptomic and proteomic data sets. 

\subsection{Allelic Influence Modules Integrate Forward and Reverse Approaches to Biomodel Engineering}

Allelic influence modules were designed to represent regulatory influences of mutated genes (the alleles of a gene) onto the system. Unlike in the other module types, these influences may be rather indirect and may involve a number of potentially unknown components. As we have shown, defining these modules allows to reverse engineer networks from data collected on mutants. These reverse engineered networks are indeed fully compatible with the molecule-centred modules through transitions that control the active states of a gene as shown in the case study. 

\subsubsection*{Acknowledgement}
We thank Mostafa Herajy, Fei Liu, Christian Rohr and Martin Schwarick for continuous support in developing Snoopy.

\bibliographystyle{splncs03}
\bibliography{lit}

\end{document}